\documentclass[%
 reprint,
superscriptaddress,
 amsmath,amssymb,
 aps,
 prl
]{revtex4-2}

\usepackage[english]{babel}

\usepackage[utf8]{inputenc}

\usepackage{graphicx}

\usepackage{bm}

\usepackage{amsthm}

\newtheorem{theorem}{Theorem}
\newtheorem{lemma}[theorem]{Lemma}
\theoremstyle{definition}

\begin{document}


\title{Optimizing quantum measurements by partitioning multisets of observables}
\author{O. Veltheim}
\email{otto.veltheim@helsinki.fi}
\affiliation{Department of Physics,  P.O.Box 64, FIN-00014 University of Helsinki, Finland}
\author{E. Keski-Vakkuri}
\affiliation{Department of Physics,  P.O.Box 64, FIN-00014 University of Helsinki, Finland}
\affiliation{InstituteQ, the Finnish Quantum Institute, Helsinki, Finland}
\affiliation{Helsinki Institute of Physics, P.O.Box 64, FIN-00014 University of Helsinki, Finland}

\begin{abstract}
Quantum tomography approaches typically consider a set of observables which we wish to measure, design a measurement scheme which measures each of the observables and then repeats the measurements as many times as necessary. We show that instead of considering only the simple set of observables, one should consider a multiset of the observables taking into account the required repetitions, to minimize the number of measurements. This leads to a graph theoretic multicolouring problem. We show that the multiset method offers at most quadratic improvement but it is achievable. Furthermore, despite the NP-hard optimal colouring problem, the multiset approach with greedy colouring algorithms already offers asymptotically quadratic improvement in test cases. 
\end{abstract}

\maketitle

\section{Introduction}

Quantum states generally cannot be measured without disturbing the state. Thus, estimates of expectation values of multiple observables typically require a large number of copies of the same state. In order to reduce the quantum computational cost, we usually aim to minimize the number of state preparations. Some tomography methods approach this problem by restricting the states that can be considered to, e.g., Matrix Product States \cite{Lanyon_2017}, or by restricting the observables that are measured to, e.g., $k$-local observables \cite{OverlappingTomography, Garcia-Perez}. Other notable methods include measuring in a random basis \cite{Huang}, or performing gentle measurements which disturb the original state as little as possible \cite{Aaronson}.

In our proposed method, no assumptions need to be made about the observables that we wish to measure, the state itself nor the types of available operations.
However, we make an implicit assumption that the number of observables grows at most polynomially with the system size. The reason for this is two-fold: With an exponential number of observables the existence of an efficient measurement scheme is unlikely. Furthermore, the exponential number of observables would inevitably mean an exponential classical overhead.

In this letter we represent partitioning the set of observables as a graph colouring problem. Similar reductions to graph colouring (or the equivalent problem of clique covers) have been considered especially in the context of Variational Quantum Eigensolvers in, e.g., \cite{Verteletskyi_2020,  Huggins_2021}. The multicolouring approach taken also in this paper was first considered in \cite{MSc}.

\begin{figure}[tb]
\includegraphics[width=\linewidth]{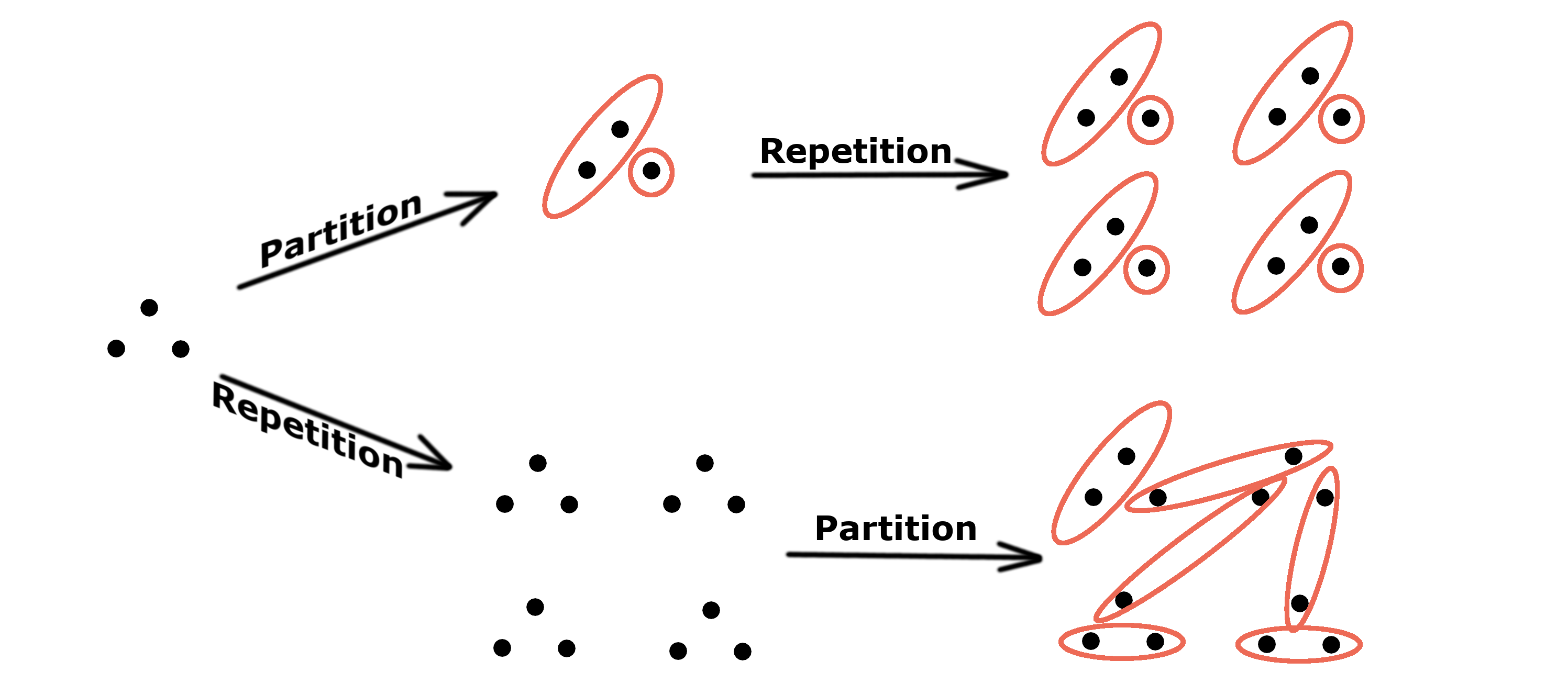}
\caption{\label{fig:Repetition-Partition} Instead of starting by partitioning the observables and then repeating measurements with the same partition, we can start by considering the number of needed repetitions and then construct the partition over the multiple instances of the same set.}
\end{figure}

A common strategy for a tomography problem is to start by trying to find as optimal partition as possible for the simple set of observables, and then repeat the measurements with that partition in order to achieve the targeted accuracy. However, we propose an alternative method taking into account the number of repetitions at the outset, as this provides a more complete and useful strategy.  Consider the following simple example illustrating the main idea. Let a set of five observables be
\begin{equation}
S=\{Z_1, Z_2, X_1Z_3, X_1 X_2, X_2X_3\},
\end{equation}
where $Z_i$ denotes the Pauli Z operator on the $i$th qubit, i.e., $Z_i=I^{\otimes i-1}\otimes Z \otimes I^{\otimes n-i}$.
Since we cannot measure non-commuting observables from the same shot, any partition of these observables must consist of at least three different sets, e.g.,
\begin{equation}
S=\{Z_1, Z_2\}\cup \{X_1Z_3, X_1X_2\}\cup\{X_2X_3\}.
\end{equation}
However, if we assume that we want to measure each observable twice, instead of repeating with the same partition, we can overlap the partitions to come up with the following, more efficient measurement strategy
\begin{eqnarray}
& (Z_1, Z_2);(X_1Z_3, X_1X_2);(X_2X_3, Z_1);\nonumber\\
& (Z_2, X_1 Z_3);(X_1X_2, X_2X_3).
\end{eqnarray}
Therefore, instead of repeating the measurements with the partition of just one set of the observables, we should use a multiset containing as many instances of each observable as the number of times we need to measure them. These two different approaches are illustrated in Fig.~\ref{fig:Repetition-Partition}. In this letter we will assume for pedagogical reasons that the multiplicity for each element is the same, but our method generalizes straightforwardly to the case of different multiplicities.

We begin by establishing a bound on the improvement that can be achieved by partitioning the multiset instead of repeating $w$ times the measurements with the partition of the simple set.
Let $d_w$ denote the number of sets required at minimum for partitioning $w$ multiples of the set of observables, which we will denote $S^w$.
\begin{theorem}
\label{th:Multiset bound}
Let $S$ be a set of observables and $S^w$ be a multiset where each element of $S$ is given a multiplicity of $w$. Also, let $d_1$ be the minimum number of sets required in partitioning $S$ into sets which can be measured from a single shot. Then, the number of sets $d_w$ required for partitioning the multiset $S^w$ is at least $\sqrt{wd_1}$. 
\end{theorem}
\begin{proof}
Each instance of an observable must belong to a different set in the partition and also the number of partitions cannot be less for the multiset than for the simple set. Therefore, we find that
\begin{equation}
d_w\geq \max(d_1, w)\geq \sqrt{wd_1}.\qedhere
\end{equation}
\end{proof}
Since $wd_1$ would be the number of shots needed for repeating the partition of the simple set, the improvement cannot be better than quadratic. However, as we will see later, this quadratic improvement in asymptotic scaling is actually achievable in some scenarios.

\section{\label{sec:Measurement}Multiset method}

We begin by defining our task as follows.
Let $S=\{O_1, O_2, \dots, O_m\}$ be a set of $m$ observables which we wish to measure. We define the following properties:
\begin{itemize}
\item A binary relation $O_i \sim O_j$ iff $O_i$ and $O_j$ can be measured in parallel.
\item $\epsilon_i\in \mathbb{R}_+$ denoting the maximum error that we allow for each observable.
\item Maximum probability of error $\delta\in [0,1]$.
For simplicity, we will in this paper divide this error probability evenly with the observables so that each observable has its error probability bounded by $\delta/m$ ensuring that the total error probability is bounded by $\delta$.
\end{itemize}
With these constraints, we wish to minimize the number of shots, i.e., the number of copies of the original system that we need.

As a warm-up, we consider first a simplified problem ignoring repetitions, i.e., trying to measure each observable once without considering the accuracies and error probabilities. We will assume that the observables $O_i$ are chosen such that we know how to measure each of them on their own from a single shot. In DV (discrete variable) quantum tomography, these would usually be Pauli strings.

There may be constraints arising from the measurement protocol that are accounted for in the binary relation $\sim$ for the observables. E.g., based on the kind of operations we are able to perform, we might be able to measure all commuting observables in parallel, or only observables which commute on every single-qubit level, meaning that the binary relation can be used to encode whether we want to use entangling measurements or only product measurements. These choices will also affect how the final measurement projectors are constructed. (We remark that $\sim$ defines a so-called independence relation, meaning that it is symmetric but not reflexive: $O_i\nsim O_i$. We can never perform multiple independent measurements of the same observable from the same shot: subsequent measurements of the same observable produce the same measurement result offering no new information. The relation is also not transitive.)

\begin{figure}[tb]
\includegraphics[width=\linewidth]{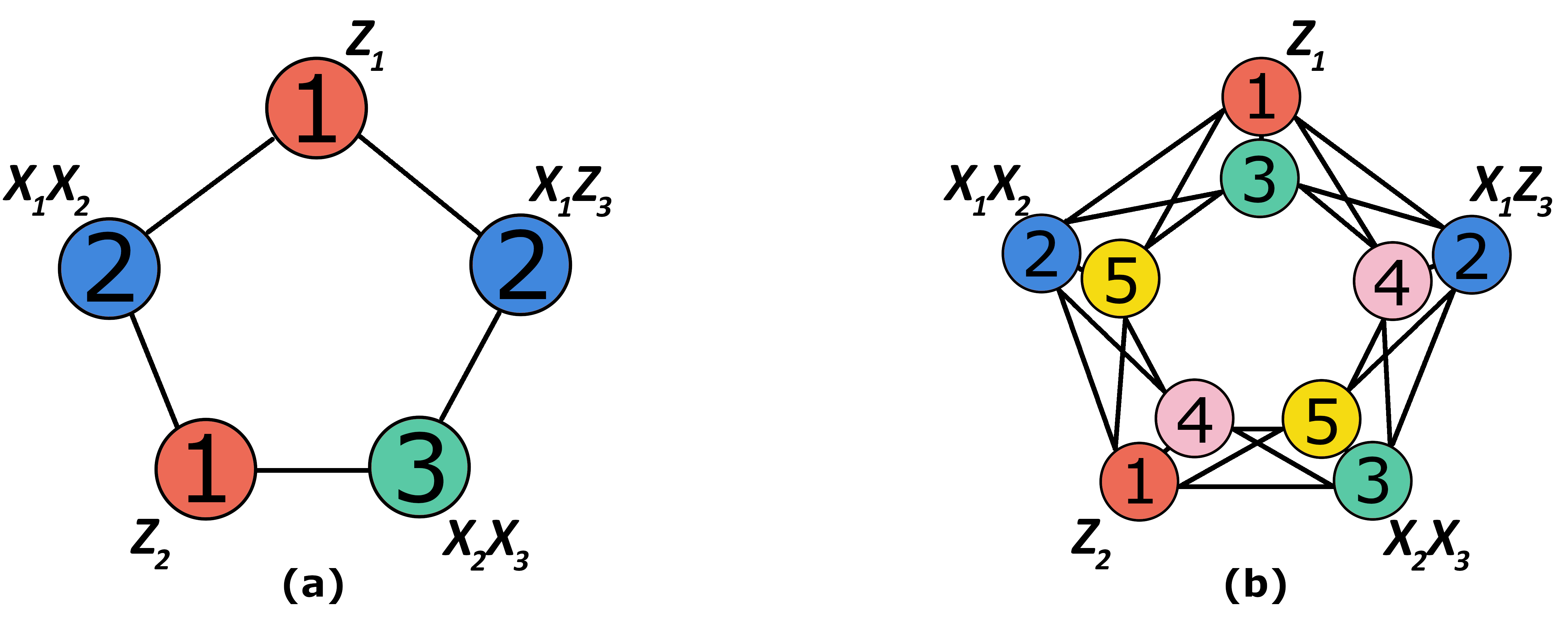}
\caption{\label{fig:Graph} (a) The commutation relations of five observables illustrated as a graph. The numbers inside the vertices denote the colour labels for graph colouring. (b) Same five observables are given two colours each corresponding to different measurements. This time five colours/shots is enough to solve the problem.}
\end{figure}

One method of representing such binary relations is with a graph. We represent each observable as a vertex of a graph and draw an edge between two vertices if the corresponding observables {\em cannot} be measured in parallel. An example of such a graph is shown in Fig.~\ref{fig:Graph}~(a).

Following these restrictions, our aim would usually be to minimize the number of shots needed for measuring all of the observables. With the graph representation, this becomes a graph colouring problem. In graph colouring, the aim is to colour each vertex of a graph with the constraint that any vertices sharing an edge cannot have the same colour. The minimum number of colours for a given graph is known as the chromatic number of the graph.

In our measurement setting, each colour in the graph colouring would correspond to needing a different copy of the system. Thus the chromatic number of the observable graph would give the minimum number of shots needed to measure every observable once. In Fig.~\ref{fig:Graph}~(a) we have coloured the observable graph with three colours giving a measurement strategy with three shots.

Solving the chromatic number of a graph and finding the corresponding colouring is an NP-hard problem so there are no known efficient, i.e., polynomial-time methods to solve them. Nonetheless, if we don't need to find the optimal colouring, i.e., the most efficient measurement scheme, there are plenty of efficient graph colouring algorithms which can find some adequate solutions most of the time (see, e.g., \cite{Lewis_2021}). As we discuss later, such colouring algorithms in the multiset method are sufficient to provide advantage.

Next, we include the repetitions in our analysis. Measuring an observable only once gives randomly one of the possible eigenvalues. In order to characterize the original system with tomography, we are more likely interested in the expectation values of the observables; thus we need to measure each observable multiple times.

Hoeffding's inequality \cite{Hoeffding} tells us that in order to achieve accuracy $\epsilon_i$ for the expectation value of observable $O_i$ with probability at least $1-\delta/m$, it is enough to measure that observable
\begin{equation}
w=\frac{2\ln \left(\frac{2m}{\delta}\right)}{\epsilon_i^2}
\end{equation}
times, where we have assumed that $O_i$ has eigenvalues $\pm 1$ like Pauli strings. Therefore, one possible and perhaps the most common solution would be to repeat $w$ times the same measurement scheme that we came up with for the single measurements. However, this will not result in the most efficient measurement strategy.

The information about measurement repetitions can be included in the set partitioning approach by using multisets, listing each observable as many times as we need to measure them. Of course, this is merely one of the many possible ways to record the information about the repetitions. Any partition of the simple set multiplied by the repetitions is a valid partition of the multiset, but the multiset also allows other partitions which are not apparent from the simple set perspective. The benefit is that the same strategies for finding those partitions work also for the multiset.

The binary relation between observables which we defined before still holds for the multiset and we can represent it as a graph as well. This is illustrated in Fig.~\ref{fig:Graph}~(b). The resulting graph replaces every vertex of the original graph with a clique of size equal to the number of repetitions. As with the simple set, colouring this graph is an equivalent problem to finding the measurement partition of the multiset. Alternatively, instead of replacing each vertex with a clique, we can also think of this as assigning each original vertex multiple colours. In that case, the problem is known in graph theory as \emph{multicolouring} \cite{Lewis_2021}.

While partitioning the multiset can never perform worse than partitioning the simple set, Theorem~\ref{th:Multiset bound} shows that the multiset approach cannot give more than quadratic improvement over the simple partition method. We will next show by an example that the quadratic improvement in asymptotic scaling can be achieved in some cases.

\section{\label{sec:RDMs}Achievability of the quadratic improvement}

To demonstrate achievability we consider an $n$-qubit system where we wish to measure every Pauli string of weight at most $k\geq 2$ (i.e., tensor products of Pauli matrices acting non-trivially on at most $k$ qubits). We will assume that we are able to perform only single-qubit operations. Even with the most naïve approaches we can measure at least $\lfloor n/k\rfloor$ of the observables from the same shot. With more elegant approaches, we can partition the set of observables into $\Theta(\log n)$ sets. We will show that this $\log n$ scaling is optimal, utilizing methods from \cite{OverlappingTomography}.
\begin{theorem}
\label{th:Partition bound}
Measuring (at least) once every Pauli string of weight at most $k\geq 2$ for an $n$-qubit system requires at least $\log_2 n$ shots.
\end{theorem}
\begin{proof}
Let us consider the problem of measuring either $Z_i X_j$ or $X_i Z_j$ for every $i,j\in \{1,\dots,n\}$. Since this set of observables is a (proper) subset of the original set of observables, this new problem clearly requires at most as many shots as the original problem. However, the minimum number of shots in this new problem is just the perfect hash family number PHFN$(n,2,2)$, which is known to be $\log_2 n$ \cite{PHF}. A more explicit proof is given in the Supplemental material S1.
\end{proof}

The above construction is also the main principle behind the overlapping tomography method in \cite{OverlappingTomography}. However, this bound was for measuring each observable only once and we need to measure each observable $\Theta\left(\frac{\log(m/\delta)}{\epsilon^2}\right)$ times. Since $m=\Theta(n^k)$, this means that we need to measure each observable $\Theta(\log n)$ times. Therefore, repeating the same measurements multiple times would result in at least $\Omega(\log^2 n)$ shots. However, partitioning the multiset will result in only $\Theta(\log n)$ required shots. 

\begin{lemma}
\label{lemma:Random}
Let us measure each qubit of an $n$-qubit system in a random Pauli basis ($X$, $Y$ or $Z$) for
\begin{equation}
d=\frac{2pw+\ln m}{p^2}
\end{equation}
shots, where $p=3^{-k}$ and $k,w,m\in \mathbb{Z}_+$. Let $s$ denote the number of times we have measured a given $k$-weight Pauli string. Then, the probability that $s\leq w$ is
\begin{equation}
P(s\leq w) < \frac{1}{m}.
\end{equation}
\end{lemma}
\begin{proof}
The probability that we measure the given Pauli string from a single shot is $p=3^{-k}$. Hoeffding's inequality tells us (see supplemental material Corollary S2.3 where, now, the random variable $X_i=1$ if we have measured the Pauli string from the $i$th shot and $X_i=0$ if we have not) that the probability of measuring the Pauli string at most $w$ times is
\begin{eqnarray}
P(s\leq w)&\leq&\exp\left[-2\left(p-\frac{w}{d}\right)^2 d\right]\nonumber\\
&=&\exp\left[-2\frac{(pw+\ln m)^2}{2pw+\ln m}\right]\nonumber\\
&< & \exp[-\ln m] =\frac{1}{m}.
\end{eqnarray}
\end{proof}
\begin{theorem}
\label{th:Multiset partition}
Let $S^w$ be a multiset consisting of every $k$-weight Pauli string of an $n$-qubit system each with multiplicity $w=a\ln n$, where $n\geq 3$ and $a\in \mathbb{R}_+$. There exists a partition of this multiset into
\begin{equation}
d=2(3^k a+3^{2k}k)\ln n
\label{eq:multisetBound}
\end{equation}
sets where any two elements of a single set commute on a single-qubit level.
\end{theorem}
\begin{proof}
Using Lemma \ref{lemma:Random} with $w=a\ln n$ and $m=\binom{n}{k}3^k$, we find that with
\begin{equation}
d=2\times 3^k a\ln n+3^{2k}\ln\left[\binom{n}{k}3^k\right]\leq 2(3^k a+3^{2k}k)\ln n
\end{equation}
shots we have measured any given $k$-weight Pauli string at least $a\ln n$ times with probability greater than $1-1/m$. Therefore, we have measured all $m$ of them at least $a\ln n$ times with probability greater than 0. This non-zero probability proves that there must exist a set of $N$ measurements which measures every $k$-weight Pauli string at least $a\ln n$ times and this set of measurements gives us the desired partition.
\end{proof}
Theorem \ref{th:Multiset partition} shows that by partitioning the multiset of observables, the required number of shots grows asymptotically only as $\Theta(\log n)$. In order to compare with the earlier lower bound for simple set partition, we should still sum eq.~\ref{eq:multisetBound} over $k$ since the simple set bound was for Pauli observables with weight \emph{up to} $k$. However, this does not affect the asymptotic scaling (as a function of $n$) so it gives a quadratic improvement to the scaling of $\Omega(\log^2 n)$ shots needed for partitioning the simple set and then repeating those measurements for the required number of times. A more explicit comparison to the overlapping tomography method is also considered in the Supplemental material S4. We will next apply
the multiset method in a concrete example, and demonstrate that a quadratic improvement can also be achieved in practice.

\section{\label{sec:Example}Example}

An important question for practical applications is the computational cost of finding the partitions as the time required for finding the optimal graph colouring scales exponentially with the graph size. Since the multiset approach also expands the size of the graph significantly, attempting to find the optimal solution would be futile. However, two simple tricks allow us to benefit from the multiset approach. First, we don't need to find the optimal solution, just some acceptably good solution. In our test cases the linear-time greedy algorithm (GA) \cite{Lewis_2021} has proven to work extremely well. GA takes the set of vertices in random order and then gives each vertex the lowest possible colour allowed by the already coloured neighbours of that vertex.

Second way to limit the computational time is to limit the number of repetitions we consider in the multiset. While the number of repetitions of measurements for each observable can be quite large, we can construct the multiset with just a fraction of those repetitions and then repeat the partition of that multiset as many times as needed. Even a small number of repeated instances of observables in the multiset can decrease the required number of shots quite significantly.

\begin{figure}[tb]
\includegraphics[width=\linewidth]{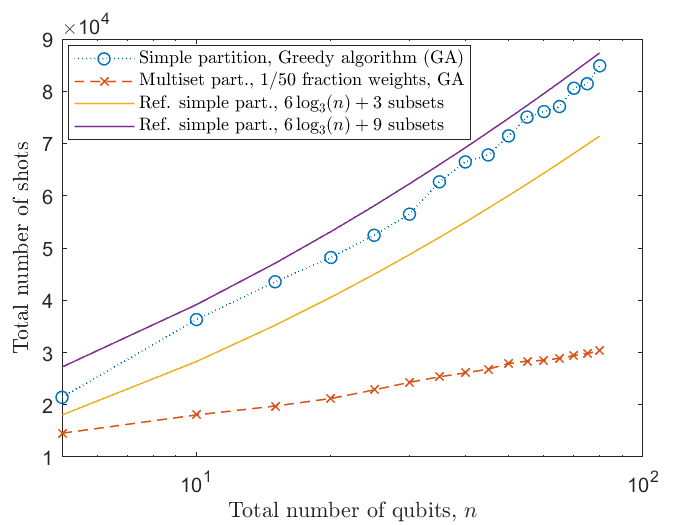}
\caption{\label{fig:Greedy} Greedy algorithm's (GA) performance for both the simple set partition and multiset partition compared to a known deterministic partition method \cite{Garcia-Perez}. GA for the simple partition (dotted line with circles), while non-deterministic, seems to perform roughly the same as the reference method (solid lines). However, GA for the multiset partition (dashed line with x's) performs much better even with just 1/50 of the actual measurement repetitions considered in the multiset.}
\end{figure}

As an example to show that the GA performs satisfactorily well, and that it is unnecessary to include the full number of measurement repetitions in the graph colouring, in Fig.~\ref{fig:Greedy} we show some experimental results for the previously considered problem of measuring $k$-local Pauli observables for the $k=2$ case with $\epsilon=\delta=0.1$. The solid lines represent some reference values for the simple partition case. More specifically, the lower solid line assumes a partition of the simple set with $6\log_3 n +3$ subsets while the upper solid line assumes a partition with $6\log_3 n +9$ subsets. These values are based on \cite{Garcia-Perez}, where it is shown that for this specific problem there exists a partition with $6\lceil \log_3 n\rceil +3$ subsets.

Our experimental GA line for the simple partition (dotted line with circles) remains within these two values, so it performs roughly the same. For the multiset partition we have limited the number of colours for each observable to 1/50 of the actual number of measurement repetitions (and then repeated that solution 50 times). With the logarithmic x-axis for the number of qubits, this multiset partition line (dashed line with x's) remains roughly linear and performs significantly better than the simple set partition lines. A pseudocode of the partition algorithm is shown in Supplemental material S6.

\section{\label{sec:Discussion}Summary and discussion}

The multiset method should always perform at least as well as any other method, especially if we compare it to a simple partition of the observables. This is because the multiset partition tries to find the optimal out of all possible mesurement strategies. As shown, the improvement over the simple partition can never be more than quadratic, but there exist problems where the asymptotic scaling achieves that quadratic improvement. We have also demonstrated methods for reducing the (classical) computational cost of partitioning the multiset by using heuristic algorithms and using only a fraction of the observable repetitions in the multiset.

We used random measurements to show the existence of a quadratically improved solution for our example problem. Therefore, it should come as no surprise that the random measurements also achieve this quadratic improvement in asymptotic scaling for this particular problem. Indeed, the methods in \cite{Huang} can be used to give an upper bound of
\begin{equation}
d=\frac{68\times 3^k\ln\left(\frac{2m}{\delta}\right)}{\epsilon^2}
\end{equation}
shots, which we have slightly improved to
\begin{equation}
d=\frac{8\times 3^{k-1}\ln\left(\frac{2m}{\delta}\right)}{\epsilon^2}
\end{equation}
for random measurements (see supplemental material). However, while random measurements would be efficient for this particular problem, they are less suited for more general problems unlike our proposed method. In the example case, we considered only Pauli observables acting nontrivially on a small number of qubits (i.e., relatively small $k$) which is particularly suited for random measurements but as soon as the observables involve a larger number of qubits, the probability of measuring them with random measurements decreases exponentially. The multiset partitioning method, on the other hand, does not have this same handicap as it treats all observables equally.

A further advantage is that this easily generalizes to problems where the needed measurement repetitions vary between observables. While most methods would end up sharing the maximum value of measurements for all observables, there is no reason why the multiplicities of each observable should be equal in the multiset. We can scale the multiplicities for each observable proportional to the number of times they need to be measured.

It would also be interesting to study whether our multiset partition approach could have applications beyond the quantum information field. The method should be suitable to any equivalent combinatorial optimization problem where a mix of independent and non-independent operations are performed repeatedly, e.g., in logistics or manufacturing.

\section*{Acknowledgements}
We thank Guillermo García-Pérez for many insightful discussions. OV's research is funded by the University of Helsinki Doctoral School. EKV acknowledges the financial support of the Research Council of Finland through the Finnish Quantum Flagship project (358878, UH).

\end{document}


\title{Supplemental material to "Optimizing quantum measurements by partitioning multisets of observables"}
\author{O. Veltheim}
\email{otto.veltheim@helsinki.fi}
\affiliation{Department of Physics,  P.O.Box 64, FIN-00014 University of Helsinki, Finland}
\author{E. Keski-Vakkuri}
\affiliation{Department of Physics,  P.O.Box 64, FIN-00014 University of Helsinki, Finland}
\affiliation{InstituteQ, the Finnish Quantum Institute, Helsinki, Finland}
\affiliation{Helsinki Institute of Physics, P.O.Box 64, FIN-00014 University of Helsinki, Finland}
\maketitle

\section{Lower bound for measuring two-qubit Pauli strings}
\begin{theorem}
Assuming only single-qubit operations and measurements, the minimum number of shots required to measure either $Z_i X_j$ or $X_i Z_j$ at least once for every pair of qubits $i,j\in\{1,\dots,n\}$ in an $n$-qubit system is $\lceil \log_2 n\rceil$.
\end{theorem}
\begin{proof}
Since we are allowed only single-qubit operations, the only way to measure an operator $Z_i X_j$ would be to measure the qubit $i$ in basis $Z$ and qubit $j$ in basis $X$ from the same shot and then multiply the two measurements together. The remaining qubits can be measured in either basis without affecting the value of $Z_i X_j$.

This means that we can represent the measurement with a binary number of length $n$. If qubit $i$ is measured in basis $Z$, then the $i$th bit in the number is set to $0$ and if it is measured in basis $X$, then the corresponding bit is set to 1. Any measurement measuring $Z_iX_j$ would then need to have $0$ in the $i$th bit and $1$ in the $j$th bit.

Now, the problem of measuring either $Z_i X_j$ or $X_i Z_j$ for every pair of qubits is equivalent to saying that in one of the binary numbers representing the measurements, the $i$th and $j$th bit must be different. We can also write the binary numbers representing measurements as an array, each row representing a different shot and each column representing different qubit. An example of this is shown in Table~\ref{tab:binary}.
\begin{table}[b]
\caption{\label{tab:binary} Measurements of a 16-qubit system with four shots. A "0" in the table means measuring the corresponding qubit in the $Z$-basis and a "1" means measuring the qubit in the $X$-basis. With just four shots, we have managed to guarantee that, for any pair of qubits, there is at least one shot where the values for that pair are not equal. This is due to the fact that we have used every four-bit number for the columns so no two columns are equal.}
\begin{ruledtabular}
\begin{tabular}{|c|cccccccccccccccc|}
\hline
& \multicolumn{16}{c|}{\textrm{Qubits}}\\
\hline
\multirow{4}{*}{\rotatebox[origin=c]{90}{Shots}}
& 0 & 0 & 0 & 0 & 0 & 0 & 0 & 0 & 1 & 1 & 1 & 1 & 1 & 1 & 1 & 1 \\
& 0 & 0 & 0 & 0 & 1 & 1 & 1 & 1 & 0 & 0 & 0 & 0 & 1 & 1 & 1 & 1 \\
& 0 & 0 & 1 & 1 & 0 & 0 & 1 & 1 & 0 & 0 & 1 & 1 & 0 & 0 & 1 & 1 \\
& 0 & 1 & 0 & 1 & 0 & 1 & 0 & 1 & 0 & 1 & 0 & 1 & 0 & 1 & 0 & 1 \\
\hline
\end{tabular}
\end{ruledtabular}
\end{table}

Furthermore, the condition that $i$th and $j$th bit must be different for at least one of the measurements is equivalent to saying that the $i$th and $j$th column in the array must be different. Therefore, measuring either $Z_i X_j$ or $X_i Z_j$ for every pair of qubits translates to filling the array in such a way that all $n$ columns are different. With $w$ shots (i.e., rows), we have $2^w$ (different) possible columns, so the condition $n<2^w$ then tells us that the number of shots we need is
\begin{equation}
w=\lceil \log_2 n\rceil.
\end{equation} 
\end{proof}

\begin{corollary}
Assuming only single-qubit operations and measurements, measuring every Pauli string of weight 2 at least once requires $\Theta(\log n)$ shots.
\end{corollary}

\section{Hoeffding's inequality}

\begin{theorem}[Hoeffding's inequality \cite{Hoeffding}]
Let $X_1,X_2,\dots, X_n$ be a set of i.i.d. random variables taking values between $a\leq X_i\leq b$. Then, the probability that their empirical mean $\mu=\frac{1}{n}\sum_{i=1}^n X_i$ exceeds the expectation value $E(X_i)$ by more than some value $\epsilon$ can be bounded by
\begin{equation}
P(\mu-E(X_i)\geq \epsilon)\leq \exp\left(-\frac{2n\epsilon^2}{(b-a)^2}\right).
\end{equation}
\end{theorem}
\begin{corollary}
In order to achieve accuracy $\epsilon$ for the mean with probability at least $1-\delta$, it is enough to measure
\begin{equation}
n=\frac{\ln\left(\frac{2}{\delta}\right)}{2\epsilon^2}(b-a)^2
\end{equation}
of the i.i.d. random variables.
\end{corollary}
\begin{corollary}
Let $X_1,X_2,\dots,X_n$ be random variables taking value $1$ with probability $p$ and value $0$ with probability $1-p$. Then, the probability that we measure $1$ less than $c$ times is
\begin{equation}
P\left(\sum_i X_i \leq c\right)\leq\exp\left(-2n\left(p-\frac{c}{n}\right)^2\right)
\end{equation}
\end{corollary}
\begin{proof}
If we consider the i.i.d. random variables $n-nX_i$ and set $\epsilon=pn-c$, Hoeffding's inequality tells us
\begin{eqnarray}
&&P\left(n-\sum_i X_i-(n-pn)\geq pn-c\right)\nonumber\\
&&\leq \exp\left(-\frac{2n(pn-c)^2}{n^2}\right)
\end{eqnarray}
and, consequently,
\begin{equation}
P\left(\sum_i X_i\leq c\right)\leq\exp\left(-2\left(p-\frac{c}{n}\right)^2n\right).
\end{equation}
\end{proof}

\section{Upper bound for measuring $k$-qubit Pauli strings with random Pauli measurements}

The methods in \cite{Huang} can be used to show that by measuring all qubits of an $n$-qubit system in a random Pauli basis,
\begin{equation}
NK=\frac{34}{\epsilon^2} 3^k\times 2\ln\left(\frac{2m}{\delta}\right)=\frac{68\times 3^k\ln\left(\frac{2m}{\delta}\right)}{\epsilon^2}
\end{equation}
shots of random measurements is enough to measure $m$ observables, each of them Pauli observables acting nontrivially on $k$ qubits, with accuracy $\epsilon$ and probability $1-\delta$. Here we will show that this upper bound on needed measurements can be improved to $\frac{8\times 3^{k-1}\ln\left(\frac{2m}{\delta}\right)}{\epsilon^2}$. This improved upper bound was first shown in \cite{MSc}.

\begin{theorem}
Let $\{O_1,\dots,O_m\}$ be a set of Pauli strings, each of weight $k$, whose expectation values we wish to estimate by measuring multiple shots in random Pauli bases. Using
\begin{equation}
d=\frac{8\times 3^{k-1}\ln\left(\frac{2m}{\delta}\right)}{\epsilon^2}
\end{equation}
shots guarantees at least $\epsilon$ accuracy for all observables, with probability at least $1-\delta$.

\end{theorem}
\begin{proof}
Hoeffding's inequality \cite{Hoeffding} tells us that if we measure $j$ times an observable $O$, then the probability that our estimate of the expectation value differs from the actual value by more than $\epsilon$ can be bounded by
\begin{equation}
\tilde{P}\leq 2\exp\left(-\frac{j\epsilon^2}{2}\right).
\end{equation}

Let $p$ denote the probability that a random measurement measures $O$ and, consequently, with probability $1-p$ we get no information about $O$. Then, the probability that, after $d$ random measurements, our estimate differs from the actual value by more than $\epsilon$ can be bounded by
\begin{eqnarray}
P&\leq& 2\sum_{j=1}^d\binom{d}{j}p^j(1-p)^{d-j}\exp\left(-\frac{j\epsilon^2}{2}\right)\nonumber\\
&=&2\left[1-p+p\exp(-\epsilon^2/2)\right]^d.
\end{eqnarray}
In order to bound the total error probability by $\delta$, we want to bound the probability of each individual observable error by $\delta/m$. Therefore, we want to find $d$ such that
\begin{equation}
2\left[1-p+p\exp(-\epsilon^2/2)\right]^d=\frac{\delta}{m}.
\end{equation}
This is given by
\begin{eqnarray}
d&=&\frac{\ln\left(\frac{\delta}{2m}\right)}{\ln[1-p+p\exp(-\epsilon^2/2)]}\nonumber\\
&\leq&\frac{\ln\left(\frac{2m}{\delta}\right)}{p[1-\exp(-\epsilon^2/2)]}\nonumber\\
&\leq&\frac{\ln\left(\frac{2m}{\delta}\right)}{p(\epsilon^2/2-\epsilon^4/8)}\nonumber\\
&\leq&\frac{8\ln\left(\frac{2m}{\delta}\right)}{3p\epsilon^2}.
\end{eqnarray}
The probability that a Pauli string of weight $k$ coincides with a random Pauli basis measurement is $p=\frac{1}{3^k}$ so the number of measurements can finally be bounded by
\begin{equation}
d=\frac{8\times 3^{k-1}\ln\left(\frac{2m}{\delta}\right)}{\epsilon^2}.
\end{equation}
\end{proof}

\section{\label{sup:QOT}Comparison with Quantum Overlapping Tomography}

In the main paper we show how the multiset method can be used, among other things, to solve the same problem as Quantum Overlapping Tomography (QOT) \cite{OverlappingTomography} with improved efficiency. In this supplemental material we will compare the two methods more explicitly. In QOT the aim is to measure every $k$-qubit subsystem of an $n$-qubit system measuring single qubits in different Pauli bases. We will here use the $k=2$ cases as a comparison. Also, for simplicity, we will consider just the two-qubit Pauli strings (i.e., products of two non-trivial Paulis acting on different qubits) and ignore the one-qubit Paulis in the calculations.

\subsection{$(n=4,k=2)$ QOT problem}

Starting with a simple case of four qubits, the measurement procedure suggested by QOT would consist of 15 different measurements repeated as many times as needed. These measurements are listed in the first column of Table~\ref{tab:QOTmeasurements}, where for example the first measurement, $XXYY$, means that we measure the first two qubits in the $X$-basis and the last two in the $Y$-basis. This $XXYY$ measurement thus measures the following six two-qubit Pauli strings
\begin{equation}
X_1X_2, X_1Y_3, X_1Y_4, X_2Y_3, X_2Y_4, Y_3Y_4.
\end{equation}
This choice of measurements guarantees that each 2-qubit Pauli string is measured at least once within these 15 measurements. However, this simple partition of the 54 required observables (six ways to choose two qubits and $3^2$ different Pauli bases to choose for those two qubits) can also be solved with graph colouring just like we do in the main paper. Using the heuristic greedy algorithm, we find a solution with only 12 different measurements listed in the second column of Table~\ref{tab:QOTmeasurements}. However, as the problem size is so small, we can also search for the optimal graph colouring with only 9 different measurements, which is listed in the third column of Table~\ref{tab:QOTmeasurements}. This has to be optimal, since already for two qubits, there are nine different choices of basis.

\begin{table}[b]
\caption{\label{tab:QOTmeasurements} The suggested measurements for partitioning a simple set of observables consisting of every 2-qubit Pauli string in a 4-qubit system. QOT requires 15 different shots while using greedy algorithm to solve the graph colouring problem we find a solution with 12 shots. Due to the small problem size, we can also use better algorithms for the graph colouring to find an optimal solution with 9 shots.}
\begin{ruledtabular}
\begin{tabular}{|c|ccc|}
\hline
&QOT&Greedy algorithm& Optimal\\
\hline
1&$XXYY$&$YXXX$&$XXXY$\\
2&$XXZZ$&$XYXX$&$XYYZ$\\
3&$YYXX$&$XXYY$&$XZZX$\\
4&$YYZZ$&$YYYY$&$YXYX$\\
5&$ZZXX$&$YXZZ$&$YYZY$\\
6&$ZZYY$&$ZZXY$&$YZXZ$\\
7&$XYXY$&$ZYZX$&$ZYXX$\\
8&$XZXZ$&$ZYXZ$&$ZZYY$\\
9&$YXYX$&$YZYX$&$ZXZZ$\\
10&$YZYZ$&$ZXZY$&-\\
11&$ZXZX$&$XZZZ$&-\\
12&$ZYZY$&$ZXYZ$&-\\
13&$XXXX$&-&-\\
14&$YYYY$&-&-\\
15&$ZZZZ$&-&-\\
\hline
\end{tabular}
\end{ruledtabular}
\end{table}

Until now we have considered the simple partition while the main idea of the multiset method is partitioning the multiset which includes the repetitions. However, as the optimal solution for the simple partition already finds the theoretical minimum of nine shots, that cannot be improved further in the multiset; the minimum will always be nine times the repetitions.

\begin{figure}[tb]
\includegraphics[width=\linewidth]{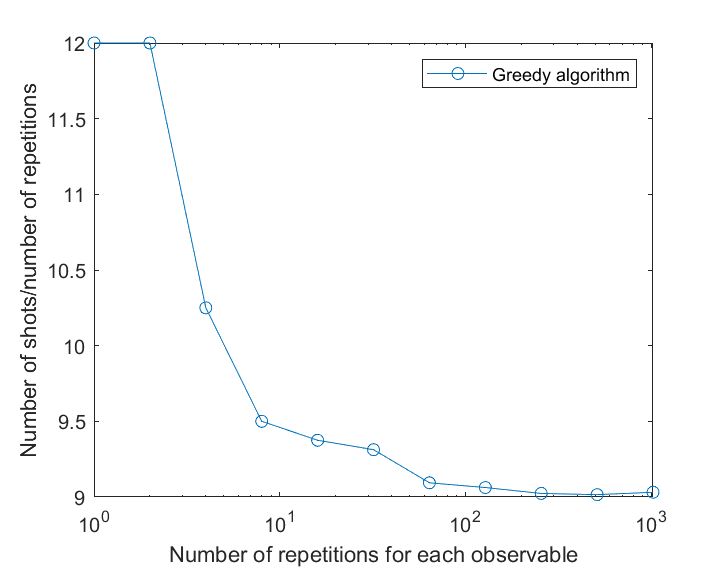}
\caption{\label{fig:GreedyReps}Greedy algorithm's performance in partitioning the multiset for the $(n=4, k=2)$ QOT problem. The number of shots divided by repetitions mapped against the repetitions themselves quickly approaches the optimal value of 9. This shows that with enough repetitions, even greedy algorithm can perform almost optimally.}
\end{figure}

Greedy algorithm, however, does improve with the repetitions. Figure~\ref{fig:GreedyReps} shows that the algorithm approaches quite fast the optimal value of nine for the number of shots divided by number of repetitions.

Clearly any single shot will always give measurements for six different two-qubit observables no matter which method we use. The difference comes from QOT dividing the measurements unevenly between the observables (18 of them are measured three times while 36 are measured only once in the simple partition), while the multiset method aims to divide the measurements as evenly as possible.

\subsection{Standard deviations in the QOT problem}

One metric to compare the different measurement protocols would be the standard deviations of observables. However, the standard deviation clearly does not depend only on the measurements and their repetitions but also the measured state. Since these methods should work for any arbitrary state, it is natural to consider the average standard deviation, where the averaging is over the relevant class of states. For a single Pauli string measurement repeated $w$ times, this average standard deviation would be
\begin{equation}
\sigma=\frac{\pi}{4\sqrt{w}}
\end{equation}
when averaged over pure states or
\begin{equation}
\sigma=\frac{9\pi}{32\sqrt{w}}
\end{equation}
when averaged over mixed states. An important thing to note here is that the standard deviation doesn't depend on the Pauli string itself (since the space of qubit states is symmetrical with respect to the different Pauli strings) or the number of qubits (since adding qubits wouldn't affect the measurement of, e.g., $Z_1$). Thus the above results can be computed with an arbitrary choice of the Pauli string. In particular, we can compute them by computing the average standard deviation of $Z$ over the Bloch sphere
\begin{eqnarray}
\sigma_\text{pure}&=&\sqrt{\frac{1}{w}}\frac{1}{4\pi}\int_0^\pi d\theta\int_0^{2\pi} \sin\theta d\varphi  \sqrt{(1-\cos^2\theta)}\nonumber\\
&=&\frac{\pi}{4\sqrt{w}}
\end{eqnarray}
or Bloch ball
\begin{widetext}
\begin{equation}
\sigma_\text{mixed}=\sqrt{\frac{1}{w}}\frac{3}{4\pi}\int_0^1 dr\int_0^\pi r d\theta  \int_0^{2\pi}r\sin\theta d\varphi  \sqrt{(1-r^2\cos^2\theta)}=\frac{9\pi}{32\sqrt{w}}.
\end{equation}
\end{widetext}
We will from here on use the average over pure states, since averaging over mixed states would just change the results by an additional constant factor.

Let us consider how many times each observable is measured in the simple partition in the QOT $(n, k=2)$ case. The total number of shots required for this simple partition in QOT is $6\log_2 n +3$. Now, let $P$ and $Q$ be some Pauli matrices from the set $\{X,Y,Z\}$ and $P_i$ and $Q_j$ are the corresponding Pauli matrices acting on the $i$th and $j$th qubit. Then, the number of times that the observable $P_iQ_j$ appears in the simple partition given by QOT is
\begin{equation}
\left\{\begin{split}2(\log_2 n-d(i,j))+1&&\text{if }P=Q\\
d(i,j)&&\text{if }P\neq Q\end{split}\right.,
\end{equation}
where $d(i,j)$ is the Hamming distance between the binary representations of $i$ and $j$. This means that the number of times each observable is measured in the single partition ranges between one and $2n-1$. Assuming that we use a total of $M$ shots to repeat the QOT protocol, the average standard deviation over all observables would be
\begin{widetext}
\begin{equation}
\label{eq:QOTSD}
\sigma_\text{ave}^\text{QOT}=\frac{1}{9\binom{n}{2}}\frac{\pi}{4}\sqrt{\frac{6\log_2 n+3}{M}}\sum_{i=1}^{\log_2 n}\frac{n}{2}\binom{\log_2 n}{i}\left[3\frac{1}{\sqrt{2(\log_2 n-i)+1}}+6\frac{1}{\sqrt{i}}\right].
\end{equation}
\end{widetext}

By comparison, we know that a theoretical optimum for the average over all observables would be when each observable would be measured from every ninth shot giving the standard deviation
\begin{equation}
\label{eq:optimalSD}
\sigma^\text{optimal}=\frac{\pi}{4}\sqrt{\frac{9}{M}}.
\end{equation}
\begin{theorem} The multiset method will reach the optimal average standard deviation if (and only if) both of the following two conditions
\begin{enumerate}
\item The number of repetitions is large enough
\item An optimal partition is found
\end{enumerate}
\end{theorem}
are satisfied.
\begin{proof}
If we choose the number of repetitions to be $3^{n-2}$ for each observable, then measuring every possible $n$-qubit Pauli string gives the optimal standard deviation.
\end{proof}

\begin{figure}[tb]
\includegraphics[width=\linewidth]{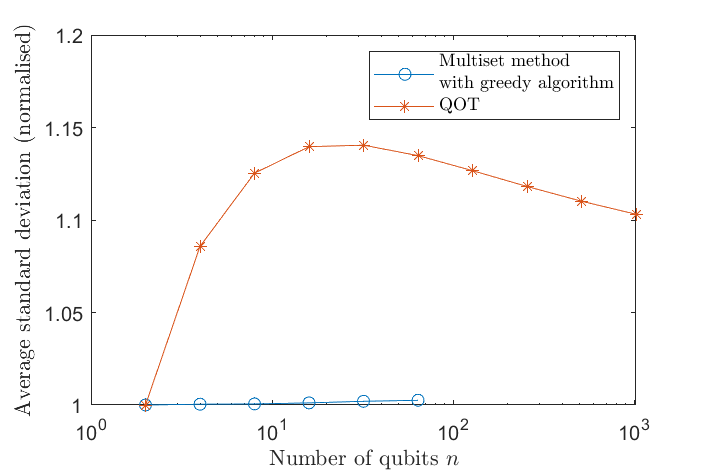}
\caption{\label{fig:SDs}Standard deviations of QOT and the multiset method averaged over all observables and normalised against a theoretical optimal value. The multiset method with greedy algorithm will stay very close to the optimal even when increasing the system size. QOT peaks at around 32 qubits at a value of around 1.14 times the optimal and then starts decreasing towards a value of approximately 1.04 times optimal. Due to computational limitations the multiset method with greedy algorithm was only simulated for up to 64 qubits while QOT values were computed analytically.}
\end{figure}

In practice, the number of repetitions is of course limited and we will not even try to find the optimal solution for larger problems. We will use Equation~\eqref{eq:optimalSD} as a reference to normalize the standard deviations. Then Equation~\eqref{eq:QOTSD} after normalization becomes
\begin{widetext}
\begin{equation}
\label{eq:normQOTSD}
\frac{\sigma_\text{ave}^\text{QOT}}{\sigma^\text{optimal}}=\frac{1}{3(n-1)}\sqrt{\frac{2\log_2 n+1}{3}}\sum_{i=1}^{\log_2 n}\binom{\log_2 n}{i}\left[\frac{1}{\sqrt{2(\log_2 n-i)+1}}+2\frac{1}{\sqrt{i}}\right].
\end{equation}
\end{widetext}
For the multiset method we will use greedy algorithm and a 1/50 fraction of repetitions needed to measure with accuracy $\epsilon=0.1$ and error probability $\delta=0.1$. As shown in Figure~\ref{fig:SDs}, the (normalised) average standard deviation will stay extremely close to the optimal value of one. However, the value for QOT also only increases to approximately 1.14 and will then asymptotically decrease towards the value 
\begin{equation}
\lim_{n\rightarrow\infty}\frac{\sigma_\text{ave}^\text{QOT}}{\sigma^\text{optimal}}=\frac{4+\sqrt{2}}{3\sqrt{3}}\approx 1.042
\end{equation}
as shown in Supplemental material~\ref{sup:QOTasymptotics}. Therefore, QOT is clearly also quite close to the optimal when comparing just the average standard deviations. However, that is not the most significant metric when comparing the methods.

As the aim is to measure all of the observables accurately, not just accurately on average, we should instead compare the worst standard deviations of all the observables. As mentioned, QOT distributes the measurements between the observables very unevenly, meaning that some observables get measured very few times compared to some others. As all of the observables should be equally important to us, we want to make sure that those observables get enough measurements as well.

\begin{figure}[tb]
\includegraphics[width=\linewidth]{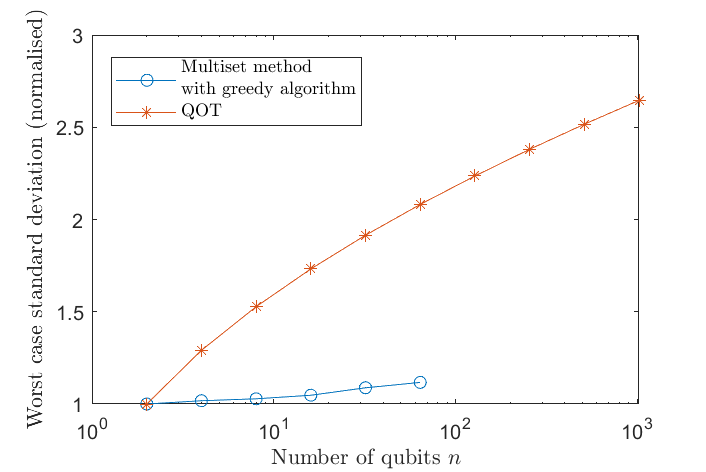}
\caption{\label{fig:worstSDs}The worst case standard deviations over all observables for QOT and the multiset method normalized against a theoretical optimal value. QOT's worst case standard deviation now clearly increases with the system size while the multiset method with the greedy algorithm manages to stay quite close tot the optimal value. Due to computational limitations the multiset method with the greedy algorithm was only simulated for up to 64 qubits while QOT values were computed analytically.}
\end{figure}

The worst case standard deviation of QOT normalised against the optimum is
\begin{equation}
\frac{\sigma_\text{worst}^\text{QOT}}{\sigma^\text{optimal}}=\sqrt{\frac{2\log_2 n+1}{3}}
\end{equation}
and we can map it again against the multiset method partitioning with the greedy algorithm, which is shown in Figure~\ref{fig:worstSDs}. Now, as we can see from the above equation, the worst case standard deviation in QOT increases as $\Omega(\sqrt{\log_2 n})\sigma^\text{optimal}$, meaning that in order to get the same accuracy, we need $\Omega(\log_2 n)$ times as many shots as an optimal multiset partitioning would.

On the other hand, as Figure~\ref{fig:worstSDs} shows, even with the greedy algorithm the multiset method manages to stay much closer to the optimal and, as proven in the main paper, with optimal partitions the multiset method would give quadratic improvement in the number of measurements needed to guarantee same worst case standard deviation.

\subsection{Error of the reduced density matrices}

As the aim of QOT is to find the $k$-qubit reduced density matrices of the system, we might also ask, how close are the measurement-based reconstructed states to the original. One way to measure this is the trace distance between the reconstructed state and the original:
\begin{equation}
    D(\tilde{\rho}, \rho)=\text{Tr}|\tilde{\rho}-\rho|,
\end{equation}
where $\rho$ is the original (reduced) state and $\tilde{\rho}$ is our estimate of the original. We can write,
\begin{equation}
    \rho=\frac{1}{2^k}\sum_{O_i\in \{I,X,Y,Z\}^{\otimes k}} \langle O_i\rangle O_i 
\end{equation}
and
\begin{equation}
    \tilde{\rho}=\frac{1}{2^k}\sum_{O_i\in \{I,X,Y,Z\}^{\otimes k}} (\langle O_i\rangle +\epsilon_i) O_i, 
\end{equation}
where $\epsilon_i$ is the error in our estimation of the expectation value of $O_i$. Now, we can write for the trace distance
\begin{eqnarray}
    D(\tilde{\rho}, \rho)&=&\text{Tr}\left|\frac{1}{2^k}\sum_{O_i\in \{I,X,Y,Z\}^{\otimes k}} \epsilon_i O_i\right|\nonumber\\
    &\leq& \sum_{O_i\in \{I,X,Y,Z\}^{\otimes k}} |\epsilon_i|\\
    &\leq& 4^k \max_i |\epsilon_i|\nonumber
\end{eqnarray}
and
\begin{eqnarray}
    D(\tilde{\rho}, \rho)&=&\text{Tr}\left|\frac{1}{2^k}\sum_{O_i\in \{I,X,Y,Z\}^{\otimes k}} \epsilon_i O_i\right|\nonumber\\
    &\geq& \sqrt{\frac{1}{2^k}\sum_{O_i\in \{I,X,Y,Z\}^{\otimes k}} \epsilon_i^2}\\
    &\geq& 2^{-k/2}\max_i |\epsilon_i|\nonumber.
\end{eqnarray}
Therefore, since
\begin{equation}
    2^{-k/2}\max_i |\epsilon_i|\leq D(\tilde{\rho}, \rho)\leq 4^k \max_i |\epsilon_i|,
\end{equation}
we know that the asymptotical scaling of trace distance with respect to $n$ is the same as the asymptotical scaling of the largest error of the expectation values so all of the previous results also hold for the trace distance.\\

\section{\label{sup:QOTasymptotics}Asymptotics of the average QOT standard deviation}

We will consider here the asymptotics of
\begin{widetext}
\begin{equation}
\frac{\sigma_\text{ave}^\text{QOT}}{\sigma^\text{optimal}}=\frac{1}{3(n-1)}\sqrt{\frac{2\log_2 n+1}{3}}\sum_{i=1}^{\log_2 n}\binom{\log_2 n}{i}\left[\frac{1}{\sqrt{2(\log_2 n-i)+1}}+2\frac{1}{\sqrt{i}}\right].
\end{equation}
\end{widetext}
which was introduced in Supplemental material~\ref{sup:QOT}. First, let us write $a=\log_2 n$, and with $a$ large, we can write the equation as
\begin{eqnarray}
\frac{\sigma_\text{ave}^\text{QOT}}{\sigma^\text{optimal}}&\approx&\frac{1}{3\times 2^a}\sqrt{\frac{2a}{3}}\sum_{i=1}^{a-1}\binom{a}{i}\left[\frac{1}{\sqrt{2i+1}}+\frac{2}{\sqrt{i}}\right]\nonumber\\
&\approx&\frac{1+2\sqrt{2}}{3\times 2^a}\sqrt{\frac{a}{3}}\sum_{i=1}^{a-1}\binom{a}{i}\frac{1}{\sqrt{i}}.
\end{eqnarray}
Now, we can use Stirling's approximation for the binomial coefficients
\begin{eqnarray}
&&\frac{\sigma_\text{ave}^\text{QOT}}{\sigma^\text{optimal}}\nonumber\\
&\approx&\frac{1+2\sqrt{2}}{3\times 2^a}\sqrt{\frac{a}{3}}\sum_{i=1}^{a-1}\frac{a^a}{i^i{(a-i)}^{a-i}}\sqrt{\frac{a}{2\pi i(a-i)}}\frac{1}{\sqrt{i}}
\end{eqnarray}
Next, we can write the summation index as $i=ax$ for some values $x\in[0,1]$ and write the summation as an integral
\begin{eqnarray}
&&\frac{\sigma_\text{ave}^\text{QOT}}{\sigma^\text{optimal}}\nonumber\\
&\approx&\frac{1+2\sqrt{2}}{3\times 2^a}\sqrt{\frac{a}{3}}\int_0^1  \frac{dx}{\left[x^x(1-x)^{1-x})\right]^a}\frac{1}{x\sqrt{2\pi (1-x)}}\nonumber\\
&\approx&\frac{1+2\sqrt{2}}{3\sqrt{6\pi}}\int_0^1 \frac{ dx\sqrt{a}}{\left[2x^x(1-x)^{1-x}\right]^a x\sqrt{1-x}}.
\end{eqnarray}
Now, noticing that for large $a$ the integrand is concentrated solely around $x=0.5$, we can first change $x\rightarrow x+0.5$ and approximate the new integrand around 0 to get
\begin{eqnarray}
\frac{\sigma_\text{ave}^\text{QOT}}{\sigma^\text{optimal}}&\approx&\frac{1+2\sqrt{2}}{3\sqrt{6\pi}}\int_{-0.5}^{0.5} dx\frac{  2\sqrt{2a}}{{(1+2x^2)}^a }.
\end{eqnarray}
Again, since this integrand is solely concentrated at $x=0$, we can as well extend the integral from $-\infty$ to $\infty$ and then use residue theorem to solve
\begin{eqnarray}
&&\frac{\sigma_\text{ave}^\text{QOT}}{\sigma^\text{optimal}}\nonumber\\
&\approx&\frac{1+2\sqrt{2}}{3\sqrt{3\pi}}\int_{-\infty}^{\infty} dx\frac{  2\sqrt{a}}{{(1+2x^2)}^a }\nonumber\\
&\approx&\frac{1+2\sqrt{2}}{3\sqrt{3\pi}}\frac{  2\sqrt{a}}{(a-1)!}\lim_{z\rightarrow \frac{i}{\sqrt{2}}}\frac{d^{a-1}}{dz^{a-1}}\frac{2\pi i}{2^a{(z+\frac{i}{\sqrt{2}})}^a }\nonumber\\
&\approx&\frac{1+2\sqrt{2}}{3\sqrt{3}}\binom{2a-2}{a-1}\frac{(-1)^{a-1}4i\sqrt{\pi a} }{2^a{(\frac{i}{\sqrt{2}}+\frac{i}{\sqrt{2}})}^{2a-1}}\nonumber\\
&\approx&\frac{1+2\sqrt{2}}{3\sqrt{3}}\binom{2a-2}{a-1}\frac{\sqrt{2\pi a} }{2^{2a-2}}.
\end{eqnarray}
Finally, using Stirling's approximation again, we find
\begin{eqnarray}
\frac{\sigma_\text{ave}^\text{QOT}}{\sigma^\text{optimal}}&\approx&\frac{1+2\sqrt{2}}{3\sqrt{3}}\frac{\sqrt{2a-2}\sqrt{a}}{a-1}\nonumber\\
&\approx&\frac{4+\sqrt{2}}{3\sqrt{3}}.
\end{eqnarray}

\section{Pseudocode for the multiset method}

A pseudocode of the multiset method algorithm can be seen in Algorithm~\ref{alg:Multicolouring}.

\begin{figure}[tb]
\begin{algorithm}[H]
\SetAlgoLined
\caption{Multiset method. Partitioning multisets of observables into measurable shots.\label{IR}}
\label{alg:Multicolouring}
\KwIn{\begin{enumerate} \item Set of observables $\{O_1, O_2,\dots, O_m\}$ \item Numbers of repetitions to consider $\{w_1, w_2,\dots, w_m\}$\end{enumerate}}
\tcc{In order to limit the computational cost, the $w_i$ can also be just fractions of the actual measurement repetitions.}
\KwOut{Partition into shots}

\tcc{Step 1: create an "adjacency list" for each observable.}
Create $m$ empty lists $\{A_1, A_2,\dots, A_m\}$.

\For{$i=1:m$}{
Add $i$ in $A_i$.

\For{$j=i+1:m$}{
\If{$O_i$ and $O_j$ cannot be measured from the same shot} {
Add $i$ in $A_j$ and $j$ in $A_i$.
}
}
}

\tcc{Step 2: Greedy algorithm to partition the observables. Can also be replaced with more sophisticated algorithms if computational time isn't an issue.}

Create $m$ empty lists $\{C_1,C_2,\dots,C_m\}$. \tcp{Lists of shots assigned for each observable}

Create a randomly ordered list $L$ of the observables with repetitions.

\For{each element $O_i$ in $L$}{
\For{$j=1:\sum w$}{
\For{each element $k$ in $A_i$}{
\If{$j$ is an element of the list $C_k$}{
Loop for next j
}
}

\tcc{None of the adjacent observables were associated with shot $j$}

Add $j$ to $C_i$.

Loop for next element in $L$.
}
}
\tcc{Now each list $C_i$ tells which shots should be used to measure observable $O_i$.}
\end{algorithm}
\end{figure}